\documentclass[%
superscriptaddress,
preprintnumbers,
 amsmath,amssymb,
 aps,
 prab,
showkeys,
twocolumn,
]{revtex4-2}

\usepackage{graphicx}
\usepackage{dcolumn}
\usepackage{bm}
\usepackage{hyperref}
\usepackage[mathlines]{lineno}
\usepackage{multirow}
\usepackage{physics}
\renewcommand{\Re}{\mathbb{R}}

\newcommand*\patchAmsMathEnvironmentForLineno[1]{%
  \expandafter\let\csname old#1\expandafter\endcsname\csname #1\endcsname
  \expandafter\let\csname oldend#1\expandafter\endcsname\csname end#1\endcsname
  \renewenvironment{#1}%
     {\linenomath\csname old#1\endcsname}%
     {\csname oldend#1\endcsname\endlinenomath}}%
\newcommand*\patchBothAmsMathEnvironmentsForLineno[1]{%
  \patchAmsMathEnvironmentForLineno{#1}%
  \patchAmsMathEnvironmentForLineno{#1*}}%
\AtBeginDocument{%
\patchBothAmsMathEnvironmentsForLineno{equation}%
\patchBothAmsMathEnvironmentsForLineno{align}%
\patchBothAmsMathEnvironmentsForLineno{flalign}%
\patchBothAmsMathEnvironmentsForLineno{alignat}%
\patchBothAmsMathEnvironmentsForLineno{gather}%
\patchBothAmsMathEnvironmentsForLineno{multline}%
}

\usepackage{academicons}
\usepackage[dvipsnames]{xcolor}
\newcommand{\orcid}[1]{\href{https://orcid.org/#1}{\includegraphics[height=3mm]{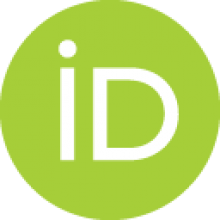}}}

\begin{document}

\preprint{APS/123-QED}

\title{Transient beam loading effects on energy loss during beam abort in high-current storage ring}

\newcommand{\qst}{\affiliation{NanoTerasu Center, National Institutes for Quantum Science and Technology (QST), Sendai, Miyagi 980-8572, Japan}}
\newcommand{\natco}{\affiliation{NAT Corporation, Hitachinaka, Ibaraki 312-0005, Japan}}


\author{Shuhei Obara~\orcid{0000-0003-3488-3553}}  \email{obara.shuhei@qst.go.jp} \qst
\author{Yuji Hosaka~\orcid{0000-0003-3084-6528}} \qst 
\author{Ryota Saida} \natco \qst 
\author{Akane Agui~\orcid{0000-0002-8299-7814}} \qst 
\author{Takao Asaka} \qst
\author{Koichi Kan~\orcid{0000-0001-6926-8468}} \qst
\author{Choji Saji~\orcid{0000-0001-9017-3906}} \qst
\author{Kota Ueshima~\orcid{0009-0005-8191-1176}} \qst
\author{Nobuyuki Nishimori~\orcid{0009-0009-1733-6292}} \qst 

\date{\today}

\begin{abstract}

Beam abort by shutting off the rf cavities is a widely used machine‑protection scheme in modern diffraction‑limited synchrotron light source storage rings. 
In this process, the stored beam loses energy turn by turn until it is intercepted by a dedicated absorber. 
A key parameter in this process is the number of turns after the rf shutdown until the subsequent beam loss, especially for the beam-size blow-up abort scheme.
Despite its importance for designing an abort protection system, this quantity has not been characterized at high stored currents.
We report measurements of abort turns over a broad current range from $3$ to $400\,\mathrm{mA}$ in the $3$‑$\mathrm{GeV}$ NanoTerasu storage ring. 
The results show a clear current dependence: the beam is lost significantly faster at higher stored currents, with the number of turns until beam loss reduced from $435$ at $3\,\mathrm{mA}$ to $187$ at $400\,\mathrm{mA}$. 
Our results indicate that transient beam loading induced by the aborting beam in empty rf cavities is the primary mechanism responsible for the enhanced energy loss. 
The number-of-turns behavior is not simply proportional to the beam loading at high current but saturates.
Time‑resolved cavity pickup signals, together with tracking simulations and analytical modeling, quantitatively reproduce the observed trend. 
Our experimental results and theoretical modeling demonstrate that transient beam loading significantly influences abort dynamics in high‑current fourth‑generation storage rings, emphasizing the need to incorporate this effect into machine‑protection system design.
Our approximate extension of the steady-state cavity–beam response to beam-abort transients reproduces experimental results well, indicating that the number of turns during a beam abort can be predicted purely numerically.
\end{abstract}
\keywords{}
                              
\maketitle

\section{Introduction} \label{sec:intro}

Beam abort systems are indispensable for machine protection in synchrotron light sources~\cite{tavares_commissioning_2018, raimondi_commissioning_2021, sirius_status_2023, apsu_tdr, thai_3gev, spring8-II}, particularly in modern diffraction‑limited storage rings that operate with ultra‑low emittance and increasingly high stored currents. 
A common abort strategy is to switch off the rf cavities~\cite{APS-U-PDR-2017, Schroer2019PETRAIVCDR, Borland2018IPACDecoherence, Borland2019NAPACBeamAborts, Dooling2019NAPACBeamDumps}, allowing the stored beam to lose energy passively until it is intercepted by a dedicated absorber. 
In this work, we refer to the number of turns from the rf‑power shutdown to the beam loss as the ``abort turns'', a key parameter that determines the effectiveness of the beam‑size blow‑up abort scheme~\cite{Hiraiwa2021-zz} using the bunch-by-bunch feedback (BBF) system~\cite{ueshima:procibic2024-tubi1}.
Because the narrow-emittance high-density beam would damage the absorber, the BBF-shaker abort system drives vertical oscillation at the resonant frequency to blow up the beam size during beam abort.
It is important to manage the vertical tune shift and the resonance time during the abort~\cite{Hiraiwa2021-zz}.
Although abort‑turn behavior has been studied in several next‑generation facilities, previous investigations have focused primarily on low‑ to moderate‑current conditions or simulation‑based studies below $100\,\mathrm{mA}$, such as those reported for APS‑U~\cite{Borland2019NAPACBeamAborts, Dooling2019NAPACBeamDumps}. 
A systematic and experimentally validated understanding of abort dynamics at several hundred milliamperes --- relevant to fourth‑generation storage rings now entering routine operation --- has been lacking. 
In particular, the influence of transient beam loading in empty rf cavities immediately after rf shutdown has not been directly quantified, despite its potential to alter beam‑energy loss and therefore the abort timeline.

In this work, we provide the first comprehensive measurement of the abort turns in the $3$‑$\mathrm{GeV}$ NanoTerasu storage ring~\cite{PhysRevAccelBeams.28.020701} over a wide stored‑current range from $3$ to $400\,\mathrm{mA}$ and clarified its behavior using our theoretical model.
We observe a pronounced current dependence: higher stored currents lead to faster beam loss, with the number of abort turns reduced by more than a factor of two between low‑ and high‑current operation. 
By combining time‑resolved cavity pickup measurements, tracking simulations, and an analytical model of transient beam loading in rf cavities, we identify this effect as the primary mechanism responsible for the accelerated energy loss during the abort.
These results provide experimental evidence that transient beam loading plays a significant role in abort dynamics in high‑current storage rings. 
As fourth‑generation light sources continue to push toward even higher stored currents, accounting for this mechanism becomes essential for designing robust and reliable machine‑protection systems.

In this paper, we present a measurement setup for beam abort responses in Sec.~\ref{sec:setup}, experimental results for the number of beam abort turns in Sec.~\ref{sec:results}, and a discussion in Sec.~\ref{sec:disc}.
Finally, the conclusion is described in Sec.~\ref{sec:conc}.

\section{Experimental Setup} \label{sec:setup}

We measured the number of turns after rf shutdown until the beam loss at the $3\,\mathrm{GeV}$ NanoTerasu storage ring. 
The accelerator has a $348.8\,\mathrm{m}$ circumference, a natural emittance of $1.14\,\mathrm{nm}\,\mathrm{rad}$, and a four-bend-achromat lattice, and can store up to $400\,\mathrm{mA}$ in 
$550\,\mathrm{bunches}$ with a harmonic number of $h=592$. 
The momentum compaction factor is $\alpha = 4.3\times10^{-4}$. 
Four TM020-mode rf cavities are installed in one of the long straight sections, providing a total accelerating voltage of $2.9\,\mathrm{MV}$ at $508.76\,\mathrm{MHz}$~\cite{PhysRevAccelBeams.28.020701}.
The shunt impedance and unloaded quality factor are $R = 6.8~\mathrm{M\Omega}$ and $Q_0 = 60\,300$, respectively, and $R/Q_0$ is $113$~\cite{EGO2024169418}.
At NanoTerasu, the coupling coefficient is adjustable by variable coupling tuners and is optimized depending on the stored beam current.

The abort beam is intercepted by the electron beam absorber (EBA)~\cite{Tamura:Pasj2021}, which is horizontally displaced by $-14\,\mathrm{mm}$ from the ideal orbit. 
The EBA is located in a dispersive section in which the horizontal 
dispersion is approximately $\eta_x \simeq 0.15\,\mathrm{m}$. 
Thus, under ideal conditions (i.e., without transient effects), the beam would reach the EBA when the relative energy loss satisfies
\begin{equation}
\frac{\Delta E_{\mathrm{loss}}}{E_0} \simeq \frac{-14\,\mathrm{mm}}{0.15\,\mathrm{m}}
  \approx -9.3\%,
\end{equation}
corresponding to energy deviation $\Delta E_{\mathrm{loss}} \approx -280\,\mathrm{MeV}$ for beam energy $E_0=3\,\mathrm{GeV}$. 
The radiation loss per turn is $U_0 = 0.62\,\mathrm{MeV}/\mathrm{turn}$ only by bending magnets~\cite{PhysRevAccelBeams.28.020701}, and its effective value including installed gap-fixed wigglers is $U_0' = 0.64\,\mathrm{MeV}/\mathrm{turn}$. 
Thus, the ideal abort turns at the NanoTerasu first operation would be
\begin{equation} \label{eq:idealturn}
t_{\mathrm{ideal}}=-\frac{\Delta E_{\mathrm{loss}}}{U_0'} \approx 435\,\text{turns},
\end{equation}
which is consistent with the measured values at low stored current~\cite{PhysRevAccelBeams.28.020701}.

Beam position monitors (BPMs) are triggered by the rf-off signal and record beam position data at $859\,\mathrm{kHz}$ in turn-by-turn (TbT) mode~\cite{ueshima:procibic2024-tubi1}.
Each dataset includes data from $3\,000\,\mathrm{turns}$ prior to the trigger, and the TbT data shown in this paper are offset by this amount. 
Beam loss timing is defined as the turn at which the BPM charge intensity drops by $90\%$, $99\%$, or $99.9\%$.
These multiple thresholds are adopted to ensure robustness over a wide current range: the $90\%$ criterion is more reliable at low current where the $99.9\%$ drop is difficult to resolve, while at high current the $99.9\%$ level better approximates the actual beam loss timing, and $99\%$ provides an intermediate reference.
Rf power was switched off under different stored-current conditions, and TbT BPM data were acquired in each measurement.

To survey the betatron-tune shift during the abort, we applied a Fourier transform to the BPM signals. 
The BBF-shaker abort system drives vertical oscillations with sufficient amplitude for tune analysis when triggered by the rf-off signal. 
For the horizontal tune measurements, we manually excite a single-pulsed large-amplitude oscillation using one of the twin kicker magnets; the resulting large beam orbit distortion triggers the orbit interlock system and induces a beam abort.

We also monitor the rf-cavity pickup voltage~\cite{Ohshima:Pasj2018}, triggered by the rf-off signal.
There is a recorded timing offset; we defined the point at which the measured pickup voltage decreases as the timing zero.
The elapsed time corresponds to the TbT BPM data with the revolution time of $T_\mathrm{rev}=1.16\,\mu\mathrm{s}$.

\section{Results} \label{sec:results}

\begin{figure}[htb]
    \centering
    \includegraphics[width=\linewidth]{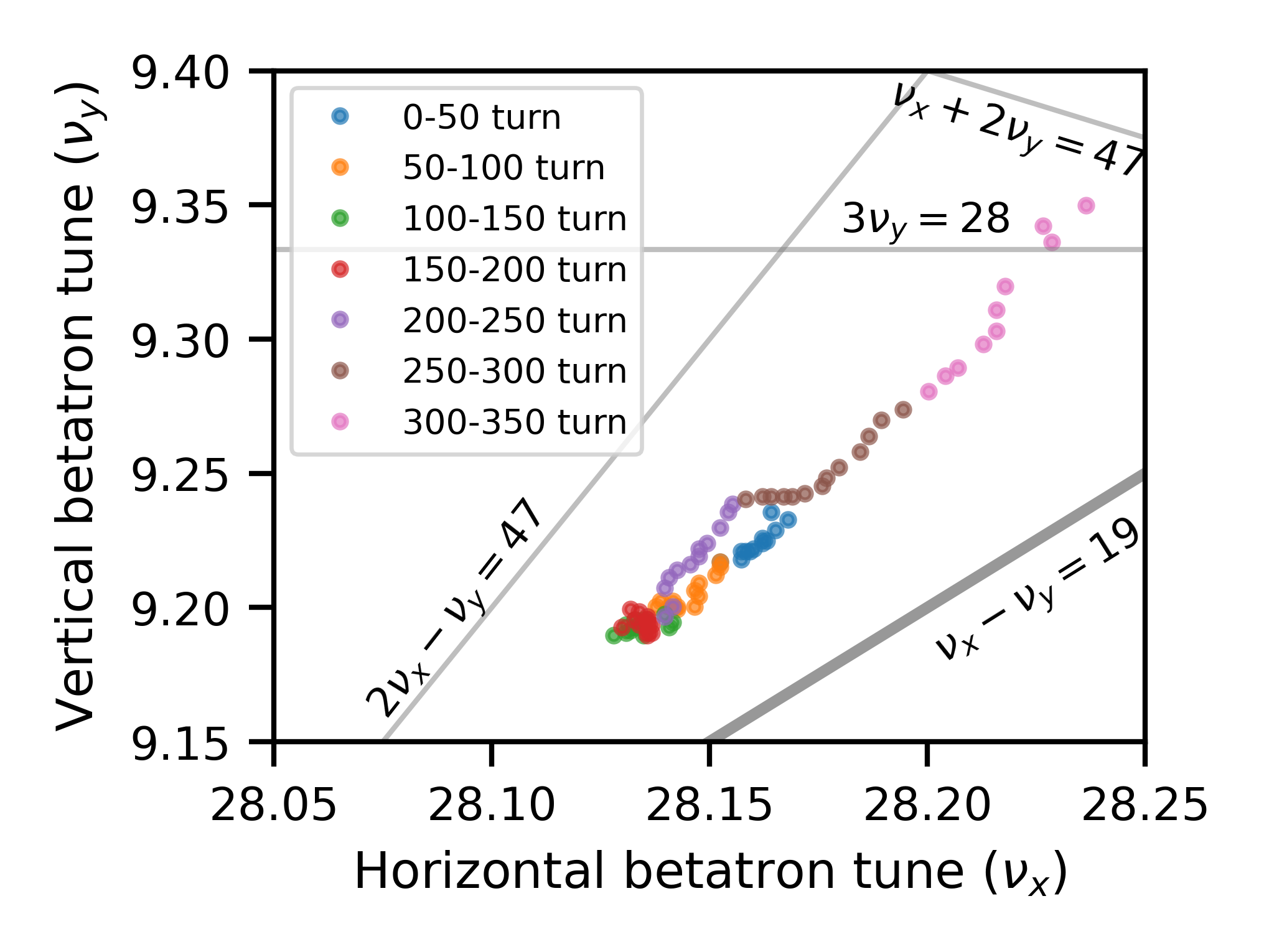}
    \caption{Betatron-tune transitions while beam abort with a $3\,\mathrm{mA}$ stored current. Gray fine lines and a bold line correspond to the third and second resonance lines, respectively.
    The operational horizontal and vertical tunes are $28.17$ and $9.23$, respectively.}
    \label{fig:plot_tune_shift_diagram}
\end{figure}

We first examined the betatron-tune shifts during the abort for a low stored current of $3\,\mathrm{mA}$, using TbT BPM data.
Figure~\ref{fig:plot_tune_shift_diagram} shows the horizontal and vertical tune transition while beam abort extracted from Fourier analysis of the BPM signals. 
The operational betatron tunes are $\nu_x = 28.17$ for the horizontal direction and $\nu_y = 9.23$ for the vertical direction.
The betatron oscillation signals remain within a narrow band during the first $\sim 350\,\mathrm{turns}$ after rf shutdown, allowing the tune values to be extracted. 
Above $\sim 350\,\mathrm{turns}$, the BPM signal becomes too weak to extract reliable tune values.
The tunes are located far from the third–order resonance lines $2\nu_x - \nu_y = 47$ and $\nu_x + 2\nu_y = 47$, so these resonances are not approached during the abort. 
Although the third–order vertical resonance $3\nu_y = 28$ is crossed during the abort tune shift, independent tune-survey measurements confirm that beam storage remains stable on this line at NanoTerasu.
The tune-survey measurements at NanoTerasu have shown vertical instability near $\nu_y = 9.5$, and a coupling resonance exists at $\nu_x - \nu_y = 19$.
The obtained data is well away from both resonances, so neither contributes to the beam loss observed during the abort.
Therefore, we conclude that the tune shift during the abort does not drive the beam loss.

\begin{figure}[htb]
    \centering
    \includegraphics[width=\linewidth]{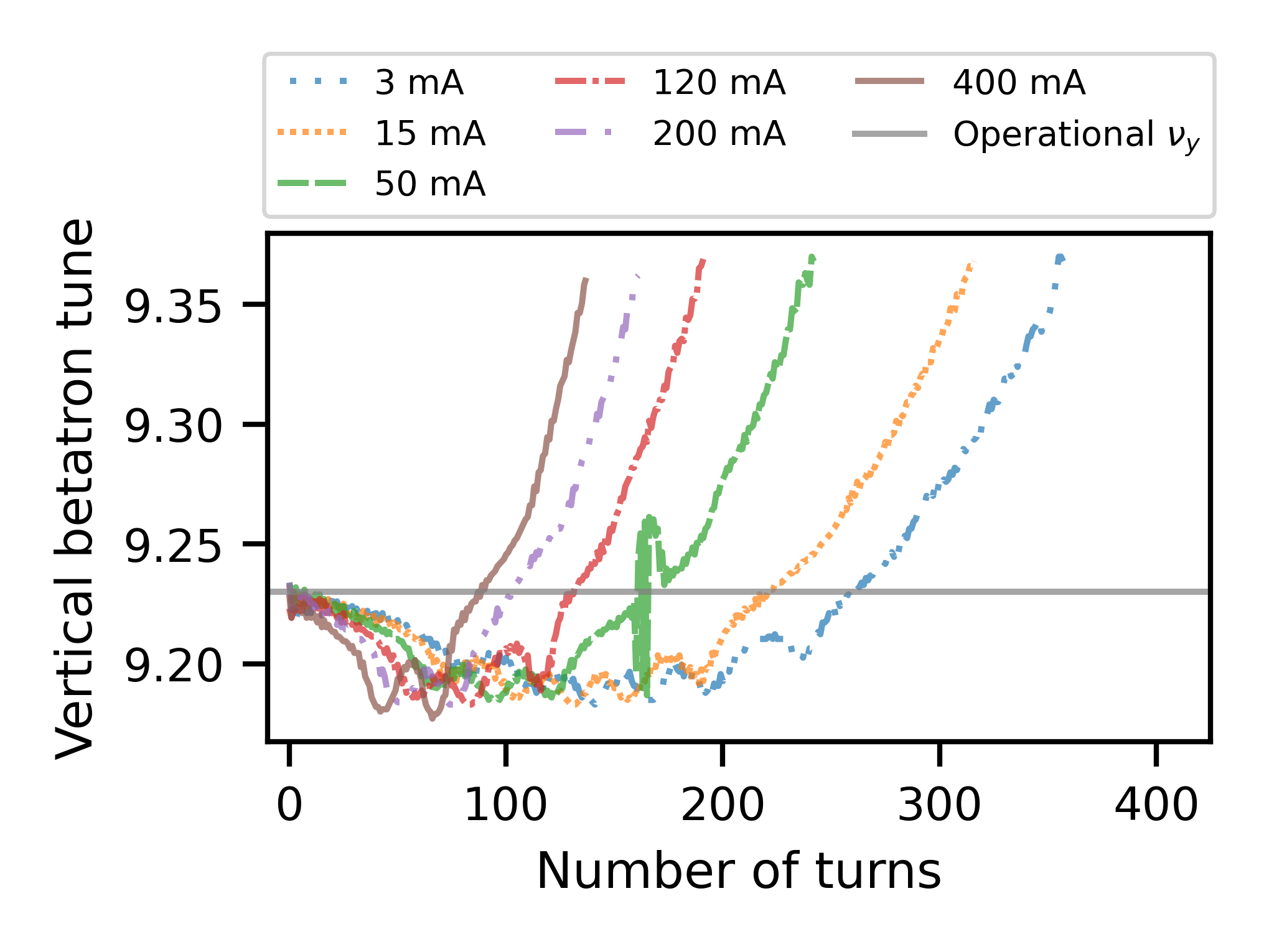}
    \caption{Dependency of vertical tune shift while beam abort on stored current.
    The horizontal gray line represents the operational betatron tune of $9.23$.
    } 
    \label{fig:plot_vtunes}
\end{figure}

To investigate the influence of the stored current on the number of turns, we measured the vertical tune shift while beam aborting at various stored-current levels.
Figure~\ref{fig:plot_vtunes} shows the vertical tune shifts as a function of the number of turns. 
Although the vertical tune follows a similar tune transition for all currents, decreasing from the operational tune of $9.23$ to $\sim9.18$ and then increasing to $\gtrapprox 9.35$, the rate of transition is significantly faster at higher currents. 
This indicates that processes increasing the energy loss --- rather than resonance crossing --- are responsible for the shortened abort duration at high stored current.

\begin{figure}[htb]
    \centering
    \includegraphics[width=\linewidth]{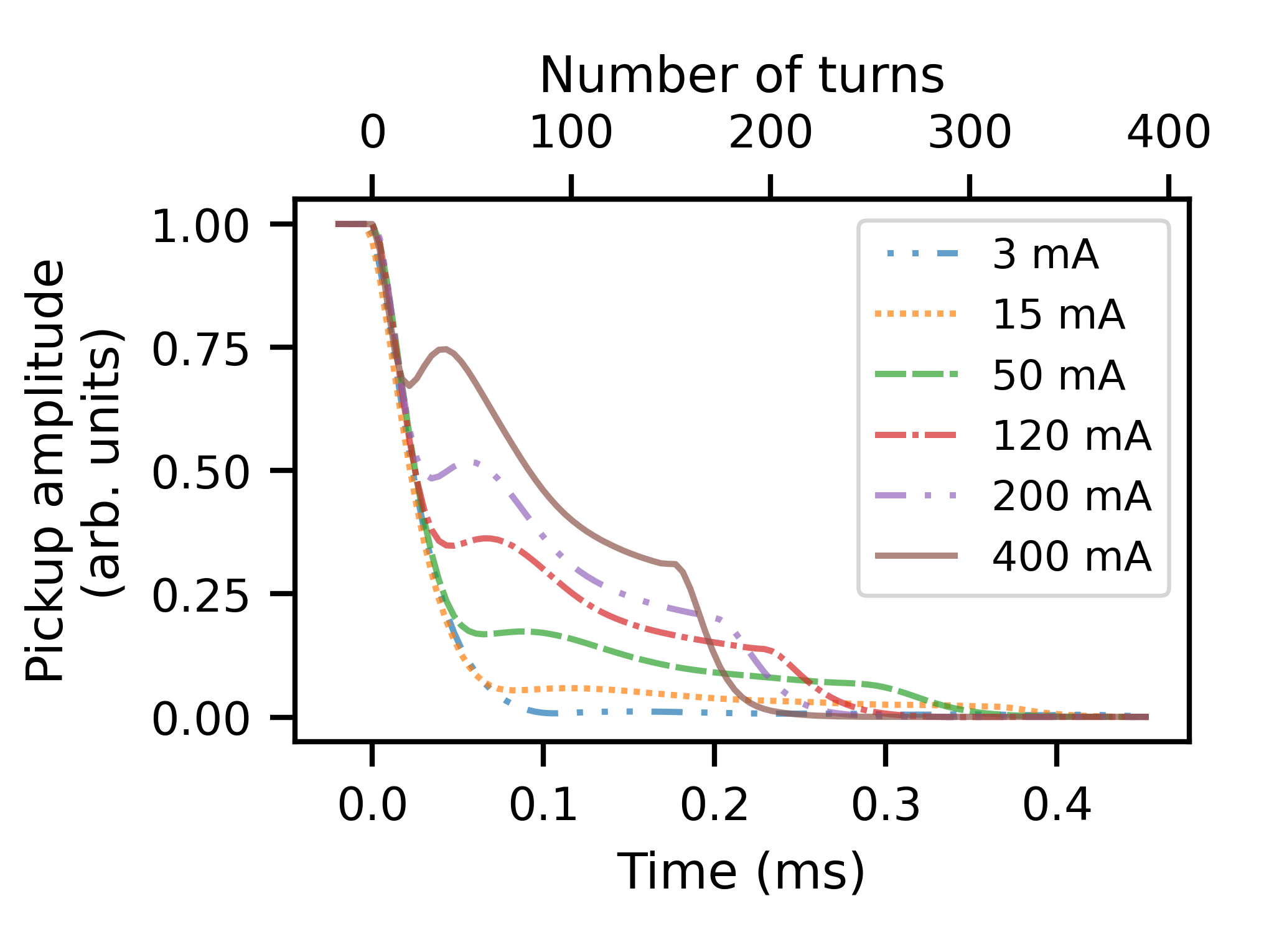}
    \caption{Dependency of normalized rf-cavity pickup voltage on stored current. 
    The vertical axis is normalized to the maximum amplitude.
    }
    \label{fig:plot_cav_pickup}
\end{figure}

We then investigated the transient field in the rf cavities during the abort by monitoring the cavity pickup voltage for various stored-current conditions.
Figure~\ref{fig:plot_cav_pickup} displays the rf cavity pickup amplitude, normalized to the rf-power-off timing. 
Even after the rf power is switched off, a beam-induced field persists in the accelerating cavities. 
The amplitude grows more strongly at higher stored currents.
This behavior indicates that the aborting beam excites a transient beam-loading voltage in the empty rf cavities, providing additional energy loss beyond the radiation loss $U_0'$.

\begin{figure}[htb]
    \centering
    \includegraphics[width=\linewidth]{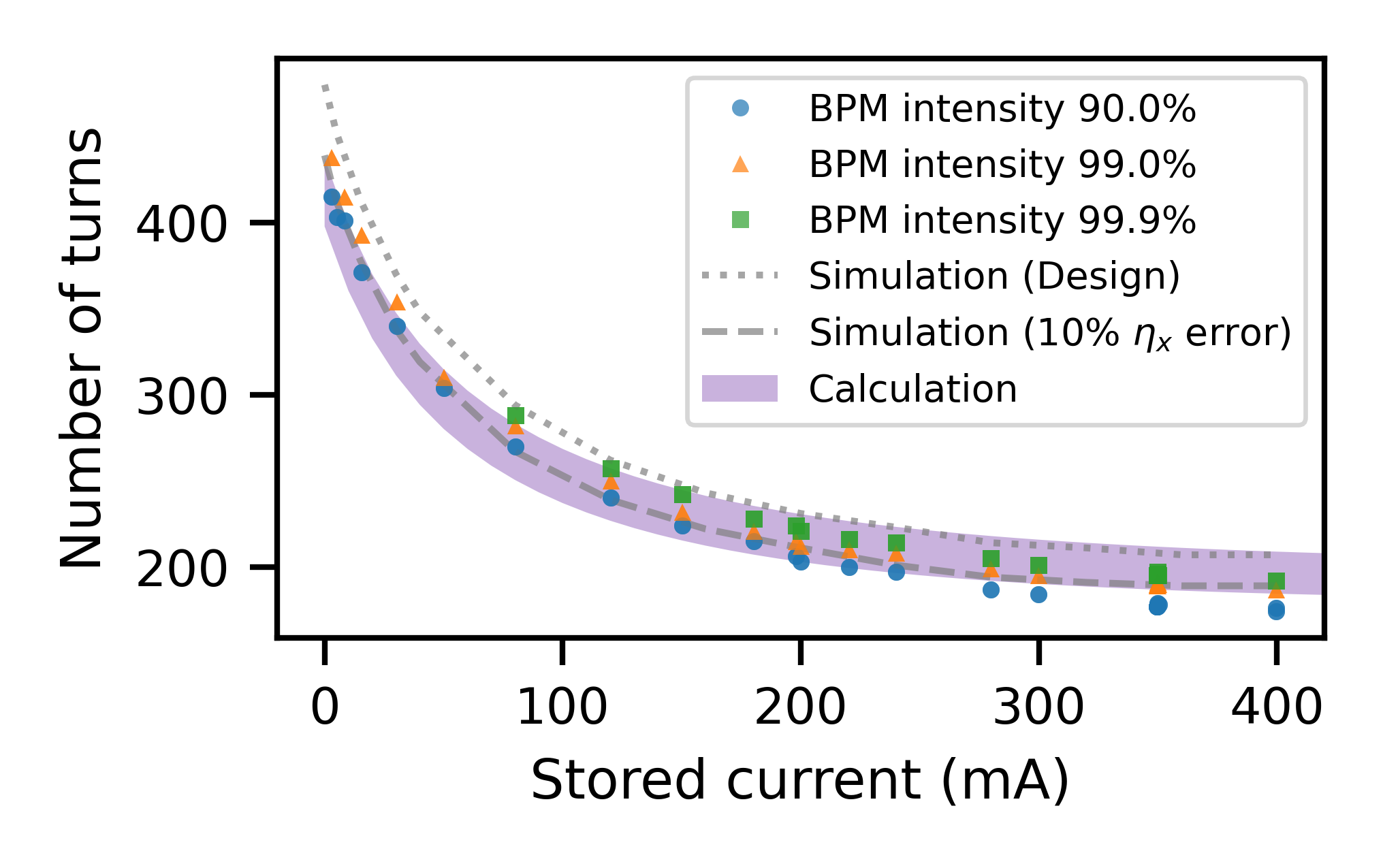}
    \caption[Dependency of abort turns vs stored current]%
    {Dependency of the number of abort turns on stored current.
    The circle, triangle, and square dots represent the timing of the listed BPM intensities lost.
    Circles indicate the $90\%$ intensity-loss criterion, triangles the $99\%$ criterion, and squares the $99.9\%$ criterion.
    Because the initial BPM total intensity is too low to calculate $99.9\%$ lost, it is shown only above $80\,\mathrm{mA}$.
    }
    \label{fig:plot_n_of_turns}
\end{figure}

Finally, we determined values of the abort turns as a function of stored current, using the TbT BPM intensity to define the beam loss time. 
We considered three thresholds for the intensity drop by $90\%$, $99\%$, and $99.9\%$. 
The resulting abort turns are summarized in Fig.~\ref{fig:plot_n_of_turns}. 
For stored currents below $\sim 80\,\mathrm{mA}$, the $99.9\%$ criterion cannot be determined because of the limited dynamic range of the BPM signal intensity; the $90\%$ and $99\%$ values remain well defined. 
The measured abort turns decrease monotonically with stored current, from $438\,\mathrm{turns}$ at $3\,\mathrm{mA}$ to $187\,\mathrm{turns}$ at $400\,\mathrm{mA}$ for data showing a $90\%$ drop in BPM intensity. 
The observed abort turns at low current agree with the ideal estimation (Eq.~\ref{eq:idealturn}).

For comparison, we show particle-tracking simulations that included the transient interaction between the aborting beam and empty rf cavities. 
The simulation reproduces the decreasing trend of the abort turns, and the simulation with $+10\%$ error in the horizontal dispersion at the EBA agrees well with the measured data. 
The dispersion uncertainty at the EBA section arises from the fact that, at present, we prioritize optimizing horizontal dispersion at the undulator section rather than the EBA section.

These results demonstrate that the stored-current dependence of the abort turns originates from the additional energy loss caused by transient beam loading in the empty rf cavities. 
A detailed interpretation of this mechanism, supported by analytical modeling, is presented in Sec.~\ref{sec:disc}.

\section{Discussion} \label{sec:disc}

Unless otherwise specified, we use the $99\%$ intensity-loss criterion as the representative definition of the abort turns, because it provides a stable signal level across the full current range.
The measurements presented in Sec.~\ref{sec:results} show that the abort turns decrease significantly with increasing stored current, from approximately $438\,\mathrm{turns}$ at $3\,\mathrm{mA}$ to $187\,\mathrm{turns}$ at $400\,\mathrm{mA}$. 
The betatron tunes remain well away from strong resonance lines during the abort process, indicating that tune shifts cannot account for the faster beam loss at high current.

Instead, the key observation is that the rf-cavity pickup power exhibits a pronounced current dependence: even after the rf power is switched off, a substantial beam-induced field builds up in the empty rf cavities at high stored current, as shown in Fig.~\ref{fig:plot_cav_pickup}. 
The cavity pickup amplitude approaches zero at $\simeq 0.10\,\mathrm{ms}$ for the low-current cases. 
After the rf shutdown, the cavity voltage decays exponentially, with the filling time defined by the loaded Q-value.
At high stored current, the pickup signal exhibits a bump around $0.1\,\mathrm{ms}$, indicating that the aborting beam induces a transient voltage that refills the cavity field even after rf shutdown.
This induced voltage affects the abort beam and accelerates its energy loss, because the beam loading voltage is in a deceleration phase in principle.
Between $0.1$ and $0.2\,\mathrm{ms}$, the induced voltage gradually decays because the beam energy loss while beam abort causes the revolution frequency change according to momentum compaction, resulting in a phase slippage with the cavity phase.
At $\simeq 0.2\,\mathrm{ms}$, the knee-like structure can be observed, which corresponds to the timing of beam loss.
When the abort beam is lost, a source of the beam-induced voltage is lost, and the cavity voltage decays.
In this context, the pickup voltage shape during the abort shown in Fig.~\ref{fig:plot_cav_pickup} is consistent with transient beam loading, and the observed bump shape can be attributed to the filling time of the beam-loading voltage and the phase-slippage effect caused by the change in the revolution time during the abort.
These features point to transient beam loading in the rf cavities as the main mechanism driving the stored-current dependence of the abort turns.

To interpret abort turns observations, we constructed a simple analytical model based on the standard beam-loading model.
Note that our analytical model uses a steady-state beam-loading model, so it takes into account the effects of phase slippage and detuning, but it does not take into account the effects of cavity filling time.
The induced cavity voltage $V_b$ is expressed as a function of the stored current $I_b$, the shunt impedance $R_{\mathrm{sh}}$, the coupling coefficient $\beta$, and the cavity tuning angle $\phi_c$:
\begin{equation}
    \Re\qty[V_b] = \frac{R_{\mathrm{sh}} I_b}{\qty{1+\beta} \qty{1+\tan^2\phi_c}},
\end{equation}
where taking the real part and the short-bunch-spacing approximation of Eq.~(\ref{eq:Vb_recursion}).
While the above equation represents the standard steady-state formalism, we now focus on a novel extension of this framework to transient, time-dependent behavior following a beam abort.
The cavity tuning angle $\phi_c$ evolves during the abort process because the relative beam-frequency deviation $\Delta\omega/\omega_c$ depends not only on the initial value of cavity detuning at the moment of rf shutdown, but also on the relative beam-energy deviation $\Delta E/E_0$ through the momentum compaction factor $\alpha$:
\begin{equation}
    \tan\phi_c(t) = 2Q_L \qty{
    \frac{\Delta\omega(0)}{\omega_c}
    -
    \alpha \frac{\Delta E(t)}{E_0}
    }.
\end{equation}
If the rf cavity is detuned to optimal frequency during operation, the initial relative beam-frequency deviation at the moment of rf shutdown $\frac{\Delta\omega(0)}{\omega_c}$ is expressed as a function of the stored current $I_b$, the shunt impedance $R_{\mathrm{sh}}$, the accelerating voltage per cavity $V_c$, the unloaded quality factor $Q_0$, and the synchronous phase $\phi_s$:
\begin{equation}
  \frac{\Delta\omega(0)}{\omega_c}
    = \frac{R_{\mathrm{sh}} I_b}{2 V_c Q_0}\cos\phi_s.
  \label{eq:delta_omega0_text}
\end{equation}
To facilitate the discussion, the dependence of the induced voltage ($\Re\qty[V_b]$) on the stored current ($I_b$) is represented as
\begin{equation}
    \Re\qty[V_b] \propto \frac{R_{\mathrm{sh}} I_b}{1+4Q_L^2\qty{\frac{R_{\mathrm{sh}}I_b}{2V_cQ_0}\cos{\phi_s} - \alpha \frac{\Delta E}{E_0}}^2}.
\end{equation}
Due to the presence of the $I_b^2$ term in the denominator of the induced voltage, rather than increasing linearly with the stored current, it exhibits saturation at higher current levels as shown in Fig.~\ref{fig:plot_n_of_turns}.
In the above expression, it is assumed that the cavity coupling parameter ($\beta$) and the loaded Q-value ($Q_L$) are constant; in NanoTerasu, they vary with the stored current, and the detailed expression can be shown in Appendix~\ref{app:beamloading}.

\begin{figure}[htb]
    \centering
    \includegraphics[width=\linewidth]{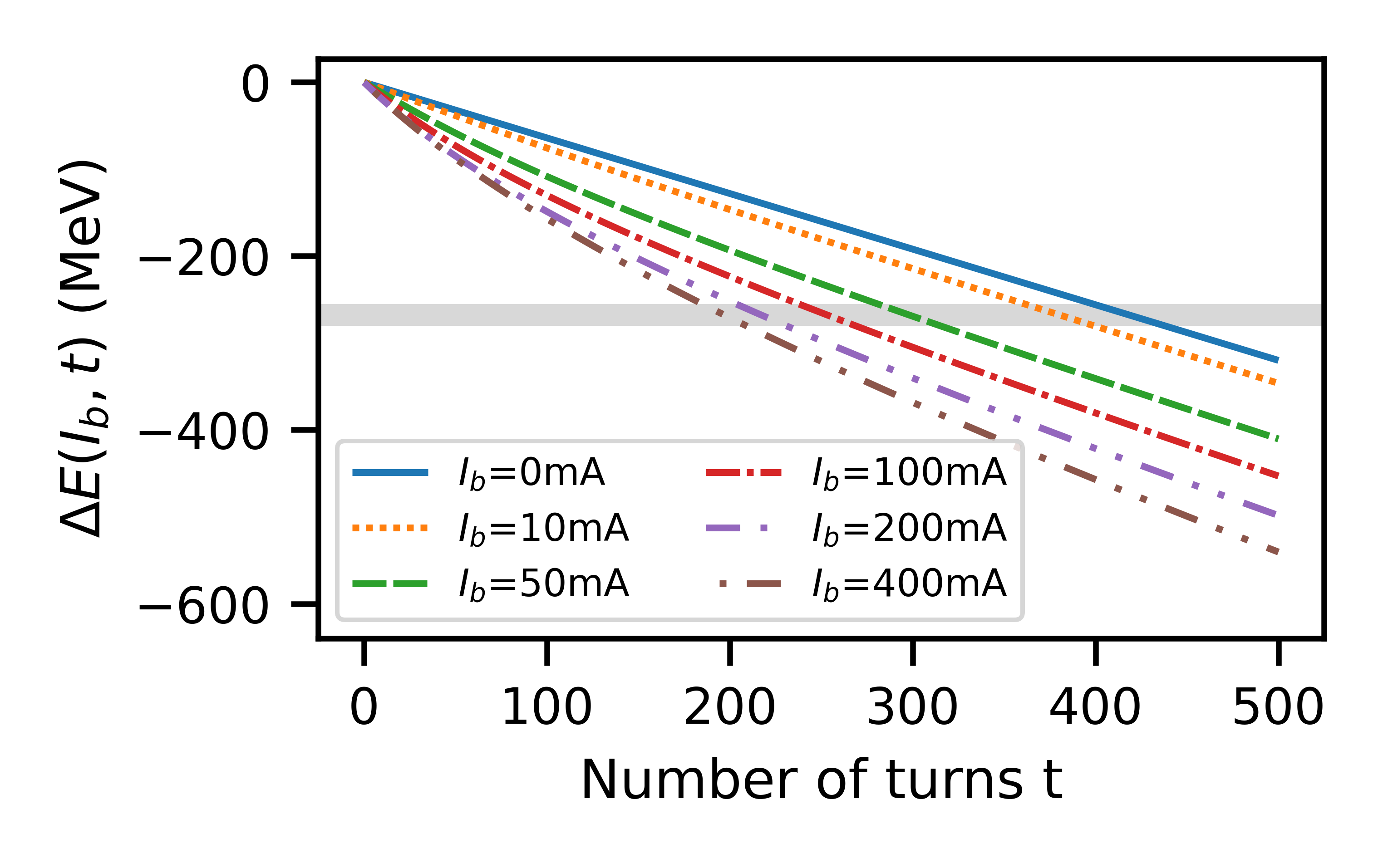}
    \caption[The total energy losses integrated for the number of turns due to the beam loading during the beam abort]%
    {The total energy losses ($\Delta E$) integrated for $t$ turns due to the beam loading during the beam abort as a function of $t$.
    The gray hatched region corresponds to $255$--$280\,\mathrm{MeV}$ of energy loss, at which the aborting beam orbit reaches the EBA. 
    This plot corresponds to a projection of the various stored currents from the Fig.~\ref{fig:dE_contour}.}
    \label{fig:rk4res}
\end{figure}

An ordinary differential equation (ODE) for the energy deviation $\Delta E(t)$ is derived, in which the energy-loss rate explicitly depends on $I_b$, $R_{\mathrm{sh}}$, $Q_0$, and $\alpha$:
\begin{equation}
    \frac{d\Delta E}{dt}
    =
    -e \Re\qty[V_b(\Delta E)] - U_0'.
\end{equation}
Although it is well-known that the beam-induced cavity voltage is proportional to the stored current $I_b$ and shunt impedance $R_{\mathrm{sh}}$, the above equations indicate that it also depends on the cavity tuning angle $\tan{\phi_c}$ and coupling coefficient $\beta$ in the denominator.
The cavity tuning angle accounts for detuning effects and beam energy losses turn by turn.
We solve the ODE numerically using a fourth-order Runge--Kutta method with the NanoTerasu cavity and lattice parameters, including the current-dependent coupling coefficient implemented via the variable coupling tuner. 
The detailed derivation of the ODE is summarized in Appendix~\ref{app:beamloading}, and the explicit form of the ODE is expressed as Eq.~(\ref{eq:dEdt_F1}).
Obtained correlation among the energy loss, number of turns, and stored current can be shown as Fig.~\ref{fig:rk4res}.

The model shows that, for NanoTerasu, the additional energy loss from transient beam loading can become comparable to the radiation loss per turn at high stored current. 
For realistic uncertainties in the horizontal dispersion at the EBA of about $+10\%$, the integrated energy loss at the abort point is $\Delta E \simeq -(255$--$280)\,\mathrm{MeV}$ (corresponding to the gray hatched band in Fig.~\ref{fig:rk4res}), which reproduces both the measured abort turns and the tracking-simulation results within the calculation error band (see Fig.~\ref{fig:plot_n_of_turns}). 
This agreement confirms that the observed reduction in abort turns with increasing stored current is predominantly driven by transient beam loading in the empty rf cavities, rather than by optics instabilities or tune-resonance effects.

As summarized in Table~\ref{tab:rfparams}, NanoTerasu combines a relatively higher shunt impedance ($R_{\mathrm{sh}} = 6.8\,\mathrm{M}\Omega$), a higher operating stored current (up to $400\,\mathrm{mA}$), and a higher overvoltage factor ($V_{\mathrm{RF}}/U_0 = 4.7$) than those of other diffraction-limited ring accelerators. 
This parameter set enhances the transient beam-loading effect, making the current dependence of the abort turns particularly pronounced in NanoTerasu.
The parameter dependence of this effect is also discussed using the analytical expressions in Appendix~\ref{app:beamloading}.
In short, increasing the transient energy loss associated with the beam loading effect $(R_{\mathrm{sh}} \times I_b)$ results in a shorter number of abort turns, whereas the overvoltage factor $(V_{\mathrm{RF}}/U_0)$ governs the saturation behavior of the number of abort turns as a function of the stored current.
\begin{table}[tb]
    \caption{\label{tab:rfparams}%
    Comparison of the shunt impedance $R_{\mathrm{sh}}$, the maximum stored current ($I_b^{\mathrm{max}}$), and the overvoltage factor ($V_{\mathrm{RF}}/U_0$).
    Here, the shunt impedance is defined as $R_{\mathrm{sh}}:=V_{\mathrm{RF}}^2/P$, where $P$ represents the total RF power loss in the cavity.
    }
    \begin{ruledtabular}
    \begin{tabular}{llll}
    \textrm{Facility}&
    $R_{\mathrm{sh}}$&
    $I_b^{\mathrm{max}}$ & $V_{\mathrm{RF}}/U_0$ 
    \\
    \colrule
    NanoTerasu & $6.8\,\mathrm{M}\Omega$ & $400\,\mathrm{mA}$ & $4.7$\\
    MAX-IV~\cite{tavares_commissioning_2018} & $\sim3.4\,\mathrm{M}\Omega$ & $500\,\mathrm{mA}$ & $2.7$\\
    APS-U~\cite{apsu_tdr} & $\sim3.8\,\mathrm{M}\Omega$ & $200\,\mathrm{mA}$ & $2.2$\\
    ESRF-EBS~\cite{esrfebs_tdr} & $\sim5\,\mathrm{M}\Omega$ & $220\,\mathrm{mA}$ & $2.6$\\
    SPring-8~\cite{inoue1991spring8rf,spring8web} & $5.5\,\mathrm{M}\Omega$ & $100\,\mathrm{mA}$ & $1.8$\\
    \end{tabular}
    \end{ruledtabular}
\end{table}

Even in accelerators where the rf coupling coefficient ($\beta$) is kept constant, the model predicts a similar trend: higher stored current leads to stronger transient beam loading and a faster reduction of the beam energy.

Owing to the optimization of the shaker frequency and shaker amplitude, the number of abort turns at $400\,\mathrm{mA}$ is still sufficient for the BBF-shaker-included abort system to blow up the vertical beam size within the first $\lesssim 50\,\mathrm{turns}$~\cite{ueshima:procibic2024-tubi1}, reducing the beam density at the EBA to a safe level. 
Thus, in the case of NanoTerasu, the transient beam-loading effect does not compromise machine protection under the present operating conditions. 
However, for future diffraction-limited storage ring accelerators aiming at even higher stored currents, detailed simulations of the aborting beam---including transient beam loading in all active and passive cavities---are essential for designing robust abort systems and dump timing.

\section{Conclusion} \label{sec:conc}

We have experimentally demonstrated that the number of abort turns---defined as the number of turns from the rf‑power shutdown to the beam loss---exhibits a clear dependence on the stored current in the $3$‑$\mathrm{GeV}$ NanoTerasu storage ring. 
Systematic measurements over a wide current range from $3$ to $400\,\mathrm{mA}$ revealed that the abort turns decrease by more than a factor of two at high stored current. 
Time‑resolved rf‑cavity pickup signals, together with tracking simulations and an analytical model, show that this behavior is primarily caused by transient beam loading in empty rf cavities, which provides additional energy loss comparable to the radiation loss per turn. 
The quantitative agreement between experiment and theory validates this interpretation.
These results provide experimental indication that transient beam loading can influence abort dynamics in a high‑current diffraction‑limited storage ring. 
As fourth‑generation light sources pursue even higher stored‑current operation, incorporating this mechanism into machine‑protection design—particularly in dump-point specification and abort‑beam‑shaping strategies—is essential for ensuring safe and reliable operation.

\begin{acknowledgments}
We thank all members of the Photon Science Innovation Center (PhoSIC) and the QST NanoTerasu Center for their help and fruitful discussions. 
QST/NAT members were dedicated to stable accelerator operations. 
The authors thank K.~Inaba (QST NanoTerasu Center) for helpful comments on an early draft of the manuscript.
We also greatly appreciate the dedication of the RIKEN SPring-8 Center, Japan Synchrotron Radiation Research Institute (JASRI), and SPring-8 Service (SES) researchers to the construction and commissioning of the NanoTerasu accelerator complex.  
The authors acknowledge Microsoft 365 Copilot for English-language editing.

\end{acknowledgments}


\appendix
\section{Analytical model of transient beam loading during rf-off abort}
\label{app:beamloading}

In this Appendix, we summarize the analytical model of transient beam loading that is used to interpret the stored-current dependence of the abort turns discussed in Sec.~\ref{sec:disc}. 
The goal is to obtain a simple expression for the additional energy loss per turn induced by the beam in the empty rf cavities after the rf power is switched off, and to derive an ordinary differential equation (ODE) for the beam-energy deviation during the abort.
The analytical expressions summarized here are standard results of beam-loading theory (see e.g. Ref.~\cite{cas2023_beamloading}), adapted to the transient rf-off abort process in NanoTerasu.

\subsection{Induced cavity voltage for a multi-bunch beam}

We consider a storage ring with rf cavities. 
The cavity is modeled as a single resonant mode with angular frequency $\omega_c$, shunt impedance $R_{\mathrm{sh}}$, unloaded quality factor $Q_0$, and coupling coefficient $\beta$. 
The corresponding loaded quality factor is
\begin{equation}
  Q_L(I_b) = \frac{Q_0}{1 + \beta(I_b)}.
  \label{eq:QL_def}
\end{equation}
The optimal value of the coupling coefficient $\beta$ depends on the stored current $I_b$.
In NanoTerasu, $\beta$ can be adjusted by a variable coupling tuner even while increasing the stored current, and the optimal $\beta$ can be written as
\begin{equation}
  \beta(I_b) = 1 + \frac{R_{\mathrm{sh}} I_b}{V_c} \sin\phi_s,
  \label{eq:beta_def}
\end{equation}
where $R_{\mathrm{sh}}$ is the shunt impedance, $V_c$ is the accelerating voltage per cavity, and $\phi_s$ is the synchronous phase at the nominal operating point.

The induced cavity voltage from a stored current $I_b$ is described by the bunch-by-bunch recursion relation
\begin{equation}
  \vec{V}_b = \vec{V}_{b0} \left\{ \frac{1}{1-e^{-\delta} e^{i\Psi}} - \frac{1}{2} \right\},
  \label{eq:Vb_recursion}
\end{equation}
where $\delta$ is the damping parameter per bunch spacing and $\Psi$ is the phase advance of the cavity field between successive bunches. 
Following the notation of the main text, we write
\begin{equation}
  \delta \equiv \delta_0 \qty{1 + \beta}, \qquad
  \Psi \equiv \delta_0 \qty{ 1 + \beta} \tan\phi_c,
  \label{eq:delta_psi_def}
\end{equation}
with $\delta_0$ being a small parameter determined by $Q_0$ and the bunch spacing, and $\phi_c$ the cavity tuning angle.

In NanoTerasu, the bunch spacing is much shorter than the cavity filling time, so that $\delta_0 \ll 1$. 
To leading order in $\delta_0$, the induced voltage can be written in the form
\begin{equation}
  V_b = I_b\, R_{\mathrm{sh}}\, \delta_0\, 
        \bigl[F_1(\beta,\phi_c) + i F_2(\beta,\phi_c)\bigr],
  \label{eq:Vb_F12}
\end{equation}
where $F_1$ and $F_2$ are dimensionless functions that represent the real and imaginary parts of the normalized beam-induced voltage, respectively.
Within this approximation, they can be expressed as
\begin{equation}
    F_1(\beta,\phi_c) \simeq \frac{1}{ \delta_0 \{1+\beta\} \{1+\tan^2\phi_c\} }, 
  \label{eq:F1_def}
\end{equation}
\begin{equation}
  F_2(\beta,\phi_c) \simeq \frac{\tan{\phi_c}}{  \delta_0 \{1+\beta\} \{1+\tan^2{\phi_c}\} },
  \label{eq:F2_def}
\end{equation}
so that $F_1$ and $F_2$ contain the full dependence on the coupling coefficient and the cavity tuning angle.

Note that the recursion relation (Eq.~\ref{eq:Vb_recursion}) assumes that bunches come constantly and does not fully reproduce transient beam loading; therefore, the delay in the rise of the induced voltage due to cavity filling time is not taken into account in this model.

\subsection{Detuning and tuning angle during the abort}
The tuning angle $\phi_c$ is related to the frequency detuning $\Delta\omega$ between the cavity resonance and the rf frequency by
\begin{equation}
  \tan\phi_c = 2 Q_L \frac{\Delta\omega}{\omega_c},
  \label{eq:phi_c_def}
\end{equation}
where $Q_L$ is given by Eq.~\ref{eq:QL_def}. 
The optimal detuning at the moment immediately before rf shutdown ($t=0$) is determined by the steady-state beam loading and can be written as
\begin{equation}
  \frac{\Delta\omega(0)}{\omega_c}
    = \frac{R_{\mathrm{sh}} I_b}{2 V_c Q_0}\cos\phi_s.
  \label{eq:delta_omega0}
\end{equation}

While the steady-state equations derived above are based on standard textbook treatments, the novel aspect of this work lies in their application to the transient dynamics associated with a beam abort.
After the rf power is switched off and the beam abort is started ($t \ge 0$), the beam loses energy and the revolution frequency shifts. 
In a synchrotron light source ring, to first order in the relative energy deviation, the detuning evolves as
\begin{equation}
  \frac{\Delta\omega(t)}{\omega_c}
    = \frac{\Delta\omega(0)}{\omega_c}
      - \alpha \frac{\Delta E(t)}{E_0},
  \label{eq:delta_omega_t}
\end{equation}
where $\alpha$ is the momentum compaction factor, $E_0$ is the nominal beam energy, and $\Delta E(t)$ is the energy difference relative to $E_0$ during the abort. 
Combining Eqs.~\ref{eq:phi_c_def} and \ref{eq:delta_omega_t}, the tuning angle can be expressed as a function of the stored current and the beam-energy deviation:
\begin{equation}
    \tan\phi_c(t) = 2 Q_L\qty{\frac{\Delta\omega(0)}{\omega_c} - \alpha \frac{\Delta E(t)}{E_0}}.
\label{eq:phi_c_Ib_t}
\end{equation}

\subsection{Energy-loss rate due to transient beam loading}
We now relate the induced cavity voltage to the additional energy loss per turn experienced by the beam. 
The real part of the induced voltage in a single cavity, $\Re\qty[V_b(t)]$, is obtained from Eq.~(\ref{eq:Vb_F12}) by inserting the time-dependent tuning angle $\phi_c(t)$ from Eq.~(\ref{eq:phi_c_Ib_t}). 
The corresponding beam-induced energy loss per turn in one cavity is
\begin{equation}
  U_b^{\rm (cav)}(t) = e\, \Re\qty[V_b(t)],
  \label{eq:Ub_cav}
\end{equation}
and, for $N_c$ identical cavities, the total beam-induced loss becomes
\begin{equation}
  U_b(t) = e\, \Re\qty[V_b(t)]\, N_c.
  \label{eq:Ub_total}
\end{equation}
During the abort, the total energy-loss rate per turn is the sum of the radiation loss $U_0$ and the beam-induced loss $U_b(t)$:
\begin{equation}
  \frac{d\Delta E(t)}{dt}
    = -\qty{U_b(t) + U_0}.
  \label{eq:dEdt_general}
\end{equation}

Equations~\ref{eq:Vb_F12}--\ref{eq:dEdt_general} define a closed ODE for $\Delta E(t)$, because $U_b(t)$ depends on $\Delta E(t)$ only through $\phi_c(t)$. 
In NanoTerasu, $\beta$ is adjusted according to the stored current by a variable coupling tuner, so writing the dependence explicitly, we obtain
\begin{widetext}
\begin{align}
    \label{eq:dEdt_F1}
    \frac{d\Delta E(t)}{dt}
    &=
    \frac{- e R_{\mathrm{sh}} I_b N_c}
    {
    \qty{ 2+ \frac{R_{\mathrm{sh}} I_b}{V_c}\sin{\phi_s} } 
    \qty[ 
    1+ 
    \qty{ \frac{2Q_0}{ 2+\frac{R_{\mathrm{sh}}I_b}{V_c}\sin{\phi_s} } }^2
    \qty{ 
    \frac{R_{\mathrm{sh}}I_b}{2V_cQ_0}\cos{\phi_s} 
    - \alpha \frac{\Delta E(t)}{E_0} 
    }^2  
    ] 
    }
    - U_0.
\end{align}
\end{widetext}

In case of the other accelerator facilities whose rf coupling $\beta$ is constant, the first factor in the denominator of Eq.~\ref{eq:dEdt_F1} ($1+\beta$) and loaded Q-value $Q_L$ become constant parameters as, 
\begin{align}
    \label{eq:dEdt_cconst}
   &\frac{d\Delta E(t)}{dt} = \notag\\
   &\frac{- e R_{\mathrm{sh}} I_b N_c}
    {
    \qty{ 1+\beta} 
    \qty[ 
    1+ 4Q_L^2\qty{\frac{R_{\mathrm{sh}}I_b}{2V_cQ_0}\cos{\phi_s} - \alpha \frac{\Delta E (t)}{E_0}}^2
    ] 
    }
    - U_0, 
\end{align}
but the stored current dependency still shows a similar trend.

\subsection{Solution and parameter dependence}

\begin{figure}[htb]
    \centering
    \includegraphics[width=1.0\linewidth]{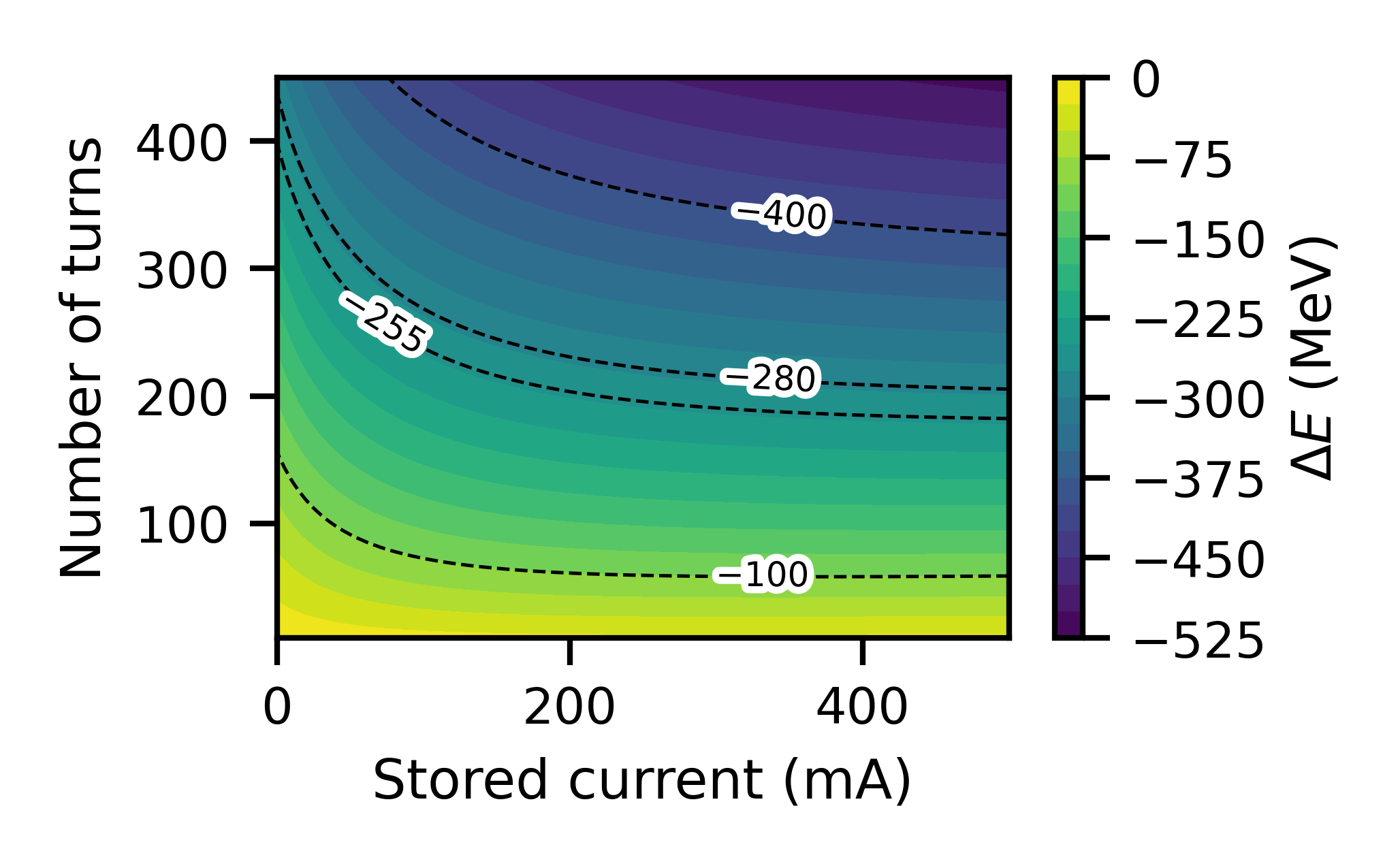}
    \caption[The total energy losses integrated for the number of turns due to the beam loading during the beam abort as a function of the stored current]%
    {The total energy losses ($\Delta E$) integrated for $t$ turns due to the beam loading during the beam abort.}
    \label{fig:dE_contour}
\end{figure}

The ODE (Eq.~\ref{eq:dEdt_F1}) is solved numerically using a classical fourth-order Runge--Kutta method (RK4) with $5000$ time steps, which provides a maximum relative residual of $\mathcal{O}(10^{-7})$ for the NanoTerasu parameter set.
Thus, with our configuration, the energy losses due to beam loading during beam abort are obtained as shown in Fig.~\ref{fig:dE_contour}. 
From $\Delta E(t)$, the number of turns required for the beam to reach the electron beam absorber (EBA) is obtained by determining the time at which the energy deviation corresponds to the horizontal displacement given by the dispersion at the EBA. 
In the main text, we consider a $+10\%$ uncertainty in the horizontal dispersion at the EBA.
Those are $\Delta E=-280\,\mathrm{MeV}$ and $\Delta E=-255\,\mathrm{MeV}$ for designed and uncertainty-included dispersion, respectively.
This translates into a band in the calculated number of abort turns, which agrees well with the experimental data and particle-tracking simulation results.

The ODE (Eq.~\ref{eq:dEdt_cconst}) can be solved analytically, although the solution is for the inverse function.
The exact solution is expressed as
\begin{widetext}
\begin{align}
    \label{eq:exact_1}
    &t(I_b, \Delta E)=-\frac{\Delta E}{U_0}\qty[ 1-\frac{E_0}{\Delta E} \frac{B I_b}{2\alpha Q_L\sqrt{B I_b+1}} \arctan \qty{\frac{\Delta E}{E_0}\frac{2\alpha Q_L\sqrt{B I_b+1}}{A^2 I_b^2-\frac{\Delta E}{E_0}2\alpha Q_L A I_b+B I_b+1}}]
    ,\\
    &A:=\frac{R_{\mathrm{sh}}}{\qty{ 1+\beta} V_c}\cos \phi_s,~B:=\frac{e N_c R_{\mathrm{sh}}}{\qty{ 1+\beta } U_0}.
\end{align}
\end{widetext}

By fixing $\Delta E$ as $\Delta E_{\mathrm{loss}}$, we can discuss the current dependence of the abort turns.
Since $-\frac{\Delta E_{\mathrm{loss}}}{U_0}$ means the ideal abort turns $t_{\mathrm{ideal}}$, the latter part of Eq.~\ref{eq:exact_1} represents the reduction in abort turns due to transient beam loading of rf cavities.
Equation~\ref{eq:exact_1} is a bit complex and unsuitable for understanding the tendency of abort turns, so using a very rough approximation, assuming that the amplitude of the arctangent is small enough and using the approximation $\arctan\theta \simeq \theta$, the abort turns can be expressed as follows:
\begin{align}    \label{eq:exact_approx}
    t(I_b)\simeq t_{\mathrm{ideal}}\qty{ 1 - \frac{B I_b}{A^2 I_b^2-\frac{\Delta E_{\mathrm{loss}}}{E_0}2\alpha Q_L A I_b+B I_b+1}}
    .
\end{align}
Equation~\ref{eq:exact_approx} can be easily differentiated with respect to stored current $I_b$: 
\begin{align}    \label{eq:exact_approx_dI}
    \frac{d t(I_b)}{d I_b}\simeq t_{\mathrm{ideal}} \frac{B \qty{A^2 I_b^2 - 1}}{\qty{ A^2 I_b^2-\frac{\Delta E_{\mathrm{loss}}}{E_0}2\alpha Q_L A I_b+B I_b+1}^2}
    ,
\end{align}
and therefore the abort turns take a minimum value around $I_b=1/A$, indicating that the abort turns at excessively high stored current may increase rather than decrease. 
Furthermore, if the relationship $I_b \ll 1/A$ is satisfied, i.e., 
\begin{align}    \label{eq:exact_approx_Ib_cond}
    \frac{R_{\mathrm{sh}} I_b}{\qty{ 1+\beta } V_c}\cos \phi_s \ll 1
\end{align}
is satisfied, the reduction of abort turns due to cavity beam loading can be ignored.
Conversely, if the relationship is not satisfied, the stored current $I_b$ and shunt impedance $R_{\mathrm{sh}}$ are sufficiently high, and the effect of transient beam loading can not be ignored.
By substituting $I_b=1/A$ into Eq.~\ref{eq:exact_approx}, we can estimate the minimum value of abort turns $t_{\mathrm{min}}$ as follows:
\begin{align}
\label{eq:exact_approx_minimum1}
    t_{\mathrm{min}} &\simeq \frac{t_{\mathrm{ideal}}}{1+C}, \\[3pt]
\label{eq:exact_approx_minimum2}
    C &:= \frac{e N_c V_c}{U_0 \cos\phi_s} \frac{1}{2-\frac{\Delta E_{\mathrm{loss}}}{E_0}2\alpha Q_L}.
\end{align}
$\frac{e N_c V_c}{U_0}$ simply represents the overvoltage factor in general, which can also be written as $1/\sin{\phi_s}$.

\begin{figure}
    \centering
    \includegraphics[width=1.0\linewidth]{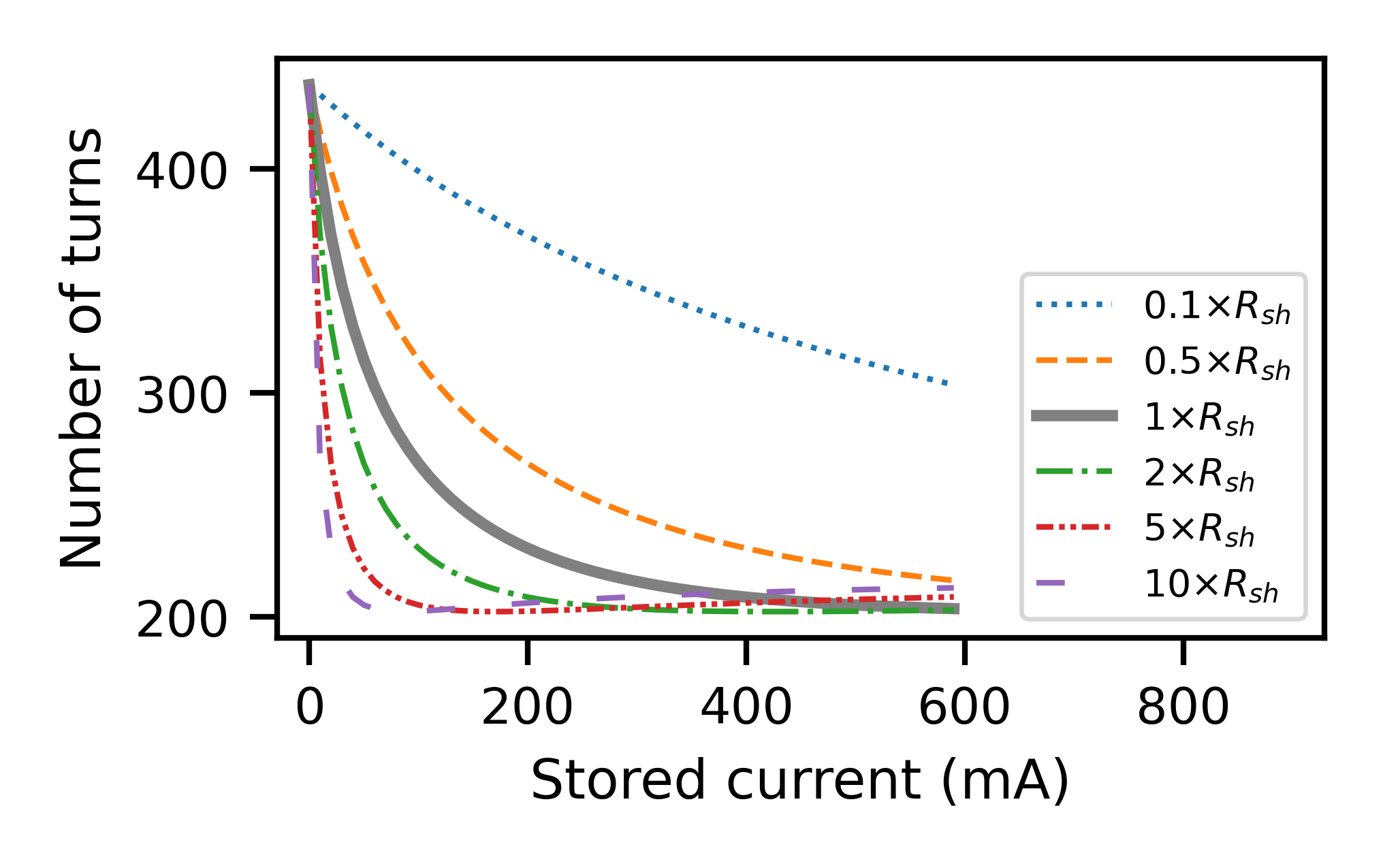}
    \caption{Dependence of the number of abort turns on the stored beam current for different values of the shunt impedance ($R_{\mathrm{sh}}$). }
    \label{fig:plot_n_of_turns_variable_shunt}
\end{figure}

\begin{figure}
    \centering
    \includegraphics[width=1.0\linewidth]{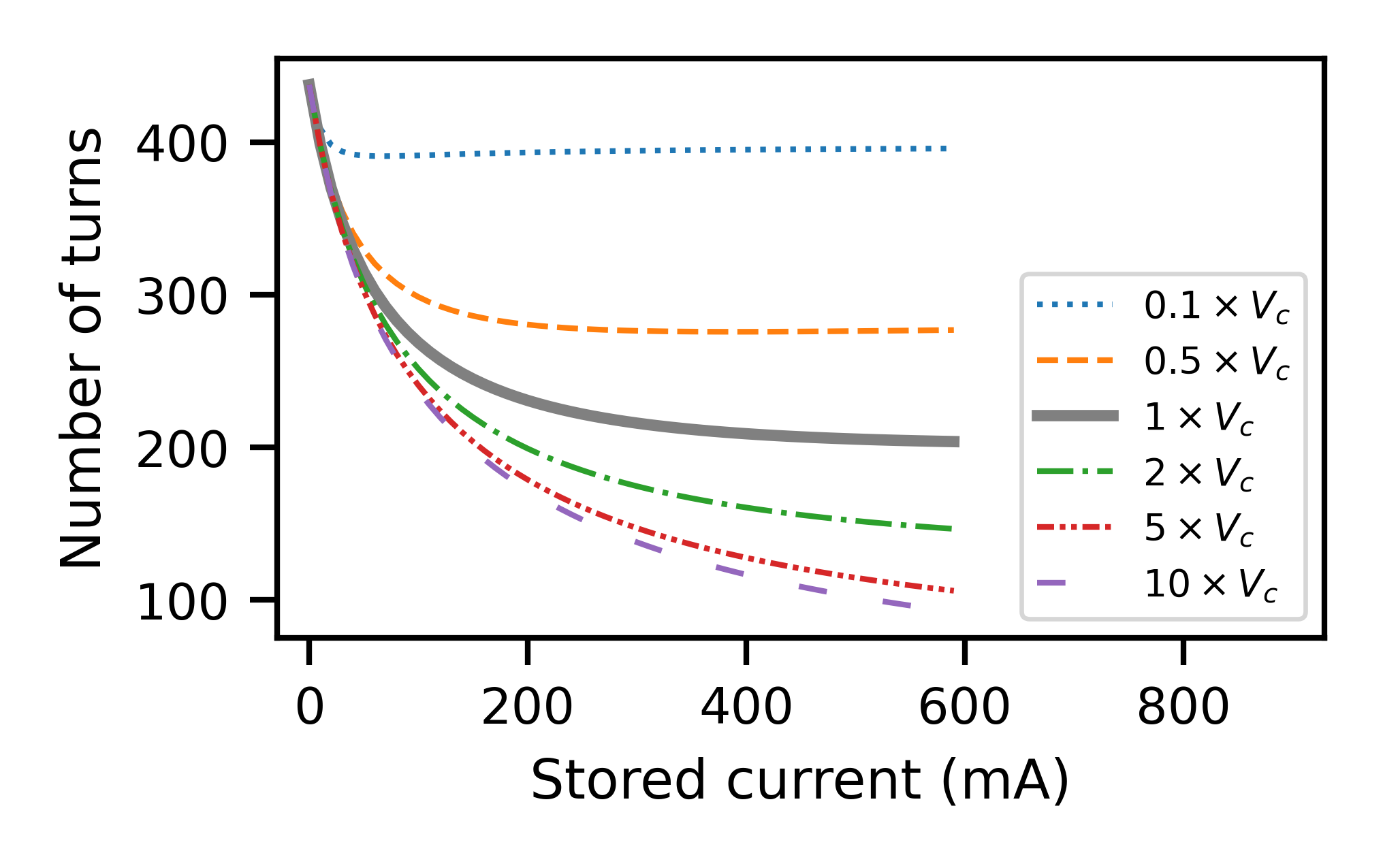}
    \caption{Dependence of the number of abort turns on the stored beam current for different values of the accelerating voltage ($V_{c}$).}
    \label{fig:plot_n_of_turns_variable_q}
\end{figure}

\begin{figure}
    \centering
    \includegraphics[width=1.0\linewidth]{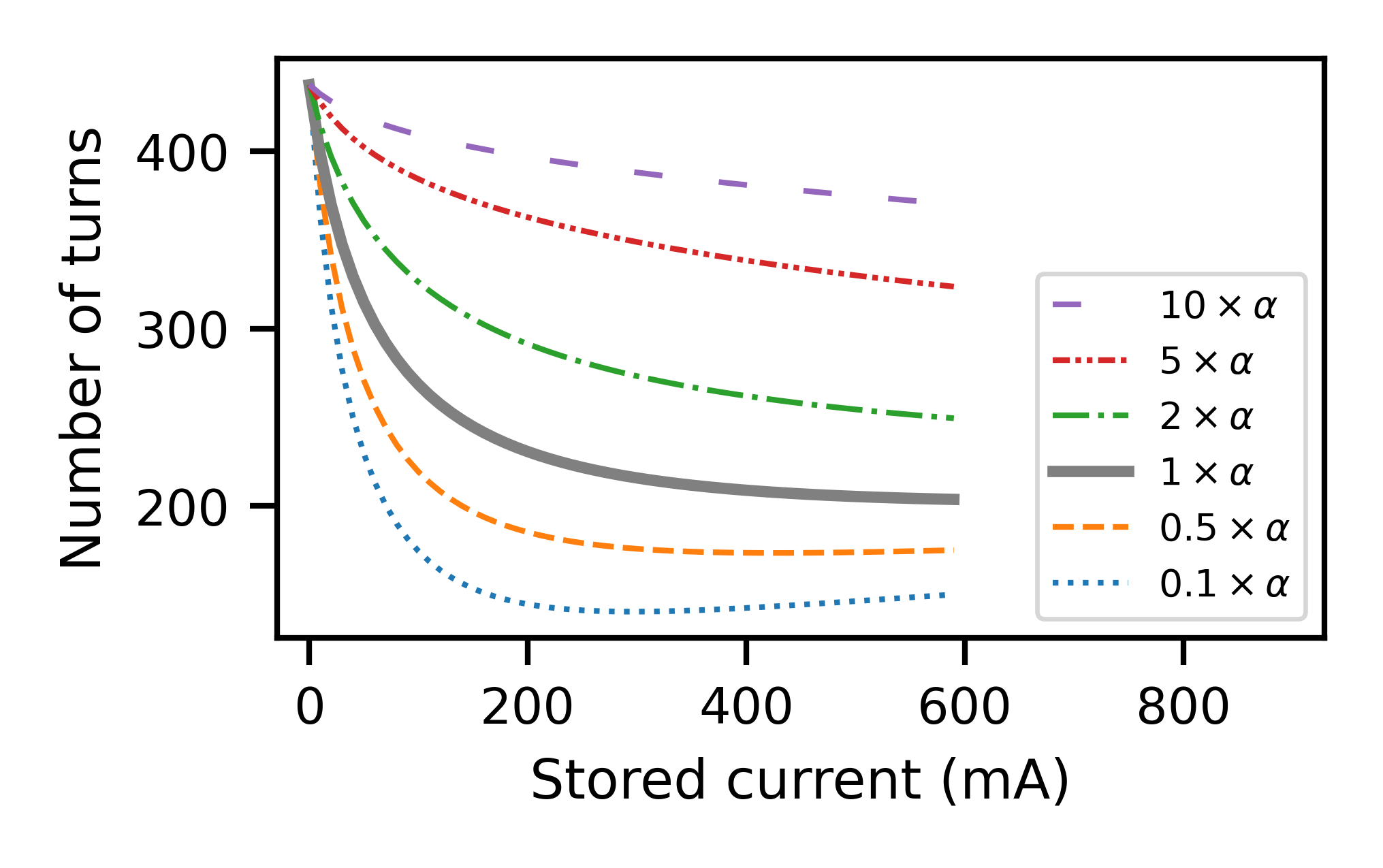}
    \caption{Dependence of the number of abort turns on the stored beam current for different values of the momentum compaction factor ($\alpha$).}
    \label{fig:plot_n_of_turns_variable_alpha}
\end{figure}

If the stored current $I_b$ and shunt impedance $R_{\mathrm{sh}}$ are high relative to the cavity voltage $V_c$, the relationship in Eq.~\ref{eq:exact_approx_Ib_cond} is not satisfied, and the reduction in abort turns due to cavity beam loading is observed.
The amplitude of the reduction in abort turns depends on the overvoltage factor as shown in Eqs.~(\ref{eq:exact_approx_minimum1},~\ref{eq:exact_approx_minimum2}), so it was strongly observed in NanoTerasu because it has high storage current, high shunt impedance, and high overvoltage factor as shown in Tab.~\ref{tab:rfparams}.
To facilitate an intuitive understanding of the response despite the complexity of the analytical expressions, we plot the dependence on the three representative parameters $R_{\mathrm{sh}}$, $V_{c}$, and $\alpha_c$, as shown in Fig.~\ref{fig:plot_n_of_turns_variable_shunt}, \ref{fig:plot_n_of_turns_variable_q}, and \ref{fig:plot_n_of_turns_variable_alpha}. 
All lines are calculation results with $\Delta E_{\mathrm{loss}} = -280\,{\mathrm{MeV}}$ condition.
As the shunt impedance ($R_{\mathrm{sh}}$) increases, the beam-loading effect proportional to ($R_{\mathrm{sh}} \times I_b$) becomes stronger, leading to a more rapid loss of beam energy and consequently a reduction in the number of turns before beam abort.
When the accelerating voltage ($V_{c}$), or equivalently the overvoltage factor, is increased, the number of turns before beam abort decreases monotonically and no longer approaches a constant value. 
In contrast, for smaller values of $V_{c}$, the number of abort turns shows only a weak dependence on the stored beam current and tends to saturate at an approximately constant value.
For smaller values of the momentum compaction factor ($\alpha$), the cavity tuning angle ($\tan{\phi_c}$) becomes less sensitive to changes in beam energy ($\Delta E$), reducing the impact of frequency detuning. 
As a result, the beam-loading term (the first term in Eq.~\ref{eq:phi_c_Ib_t}) becomes dominant, leading to a more pronounced dependence of the abort turn number on the stored beam current.



\providecommand{\noopsort}[1]{}\providecommand{\singleletter}[1]{#1}%

\end{document}